\documentclass[10pt]{aastex631}

\usepackage[utf8]{inputenc}
\usepackage{amsmath,amssymb}
\usepackage{graphicx}
\usepackage{comment}
\usepackage{svg}
\usepackage{chemformula}

\begin{document}
\title{Nucleation of water ice under weak ionization via clustering on negative ions}

\author{Andr\'e Nicolov}
\affiliation{Department of Applied Physics and Materials Science, California Institute of Technology, Pasadena, CA 91125, USA}

\author{Efstratios M. Kritikos}
\affiliation{Department of Applied Physics and Materials Science, California Institute of Technology, Pasadena, CA 91125, USA}
\affiliation{Department of Physics, University of Oxford, Oxford, OX1 3PU, England}

\author{William A. Goddard III}
\affiliation{Department of Applied Physics and Materials Science, California Institute of Technology, Pasadena, CA 91125, USA}

\author{Paul M. Bellan}
\affiliation{Department of Applied Physics and Materials Science, California Institute of Technology, Pasadena, CA 91125, USA}

\begin{abstract}
Water-ice grains are shown to form spontaneously in a cold, weakly-ionized laboratory plasma with trace water admixture when neutrals and ions are cooled below $\sim$200 K. These conditions serve as an analog of ice nucleation in cold astrophysical and atmospheric environments. The ice grains form without the presence of particulate impurities, and do not form in the absence of ionization, indicating that nucleation is induced by ions. Time-resolved ion mass spectrometry provides separate detection of clusters of water molecules on both positive and negative ions. Hydrated negative ions, mainly $\mathrm{OH^-(H_2O)_n}$, grow to $\sim$200 amu, while hydrated positive ions, mainly $\mathrm{H^+(H_2O)_n}$, remain small and short-lived. As the temperature is reduced from room temperature toward 200 K, negative ion clusters grow in size and then drop sharply in abundance, consistent with the onset of rapid growth into macroscopic ice grains below a critical temperature. This indicates that hydrated negative ions are key precursors to ice nucleation in cold, weakly ionized environments, such as the Earth's mesosphere and protoplanetary disks, providing a concrete pathway for ice formation in environments where the nucleation mechanism has remained unresolved. This negative ion pathway is estimated to dominate over ice deposition on pre-existing dust grains in protoplanetary disk locations where there is sufficient UV to partially dissociate water molecules, and in addition, where negative ions such as OH- can form readily through electron attachment at low temperature.
\end{abstract}


\section{Introduction}
Water ice is among the most abundant solids in the universe, yet the molecular pathway by which ice nucleates from vapor remains unresolved. It is conventionally assumed that the ice nucleates heterogeneously, that is, water molecules condense on pre-existing ``seed" particulates composed of other, non-water molecular species  \citep{gumbel2009charged, liu2022free}. Homogeneous nucleation from vapor is generally energetically unfavorable, yet pure water-ice aerosols are frequently observed in cold, weakly ionized environments, from noctilucent clouds in the Earth's polar mesosphere to the outer icy reaches of protoplanetary disks and interstellar clouds, without clear signatures of foreign seed particles \citep{witt1969nature, yu2001molecular, plane2018impacts, potapov2021dust}. 

Ion-induced nucleation has long been proposed for aerosol formation in weakly ionized environments, though the precise molecular pathway has not been fully characterized and its overall contribution to ice formation has often been overlooked \citep{arnold1980ion, svensmark2007experimental, gumbel2009charged, bellan2022}. Furthermore, experiments have long shown enhanced water aerosol production in the presence of ions, as ion-dipole interactions stabilize small water clusters and reduce the nucleation energy barrier \citep{wilson1897, yu2001molecular, arnold1980ion, froyd2003experimental, bryantsev_Goddard2009evaluation, guiu2025oxygen}. However, direct observation of the growth of these clusters into macroscopic particles has remained elusive, leaving a persistent size gap between sub-nanometer hydrated ions and observable ice particulates \citep{yu2001molecular}. Most laboratory studies also operate near atmospheric pressure, whereas astrophysical and upper-atmospheric ice forms at pressures well below the water triple point where the absence of a stable liquid phase makes nucleation strongly temperature-dependent \citep{witt1969nature, gumbel2009charged, draine2011physics}.

\begin{figure}[t]
    \centering
    \includegraphics[width=\linewidth]{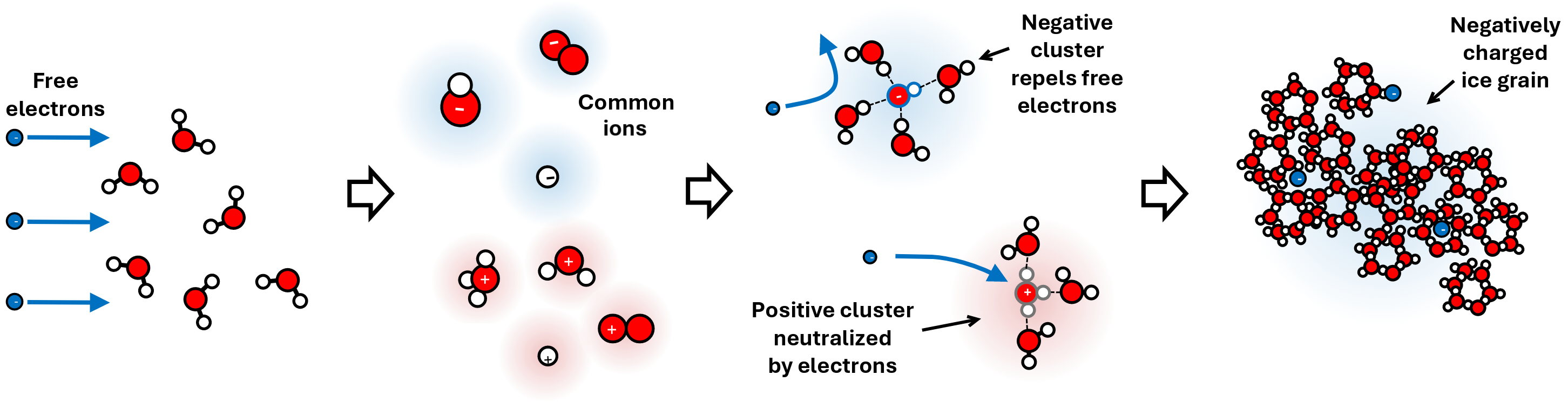}
    \caption{Free electrons collide with water molecules to produce positive and negative ions, predominantly $\mathrm{OH^-}$ and $\mathrm{H_3O^+}$; water molecules cluster around the ions; positive clusters are neutralized by the electrons, while negative clusters grow into stable ice grains. \citet{cottin1959etude} shows that the electron energy for producing $\mathrm{OH^-}$ is about 4.5 eV whereas the electron energy for producing $\mathrm{H_3O^+}$ is about 12 eV. }
    \label{fig:nucleation-sketch}
\end{figure}

We report laboratory experiments and associated theoretical models showing that ice nucleates spontaneously in cold, low-pressure, weakly-ionized conditions via rapid hydration of negative ions. The experiments use a temperature-controlled, low-pressure $\mathrm{Ar}$ -- $\mathrm{H_2O}$ gas discharge in which the clustering of water molecules around ions is measured through mass spectrometry and the associated growth of ice particles to visible sizes is tracked optically. Clusters of water molecules on negative ions abruptly disappear below a critical temperature, whereas clusters of water molecules on positive ions remain essentially unchanged. This distinction indicates that $\mathrm{H_2O}$ clusters on  negative ions  serve as the primary nucleation sites and are thus consumed in the synthesis of aerosols, while $\mathrm{H_2O}$ clusters on positive ions are rapidly neutralized by ambient electrons and so fragment faster than they can grow to become macroscopic grains. Figure \ref{fig:nucleation-sketch} summarizes this nucleation pathway.

\section{Experiment and results}
\begin{figure}[h!]
    \centering
    \includegraphics[width=\linewidth]{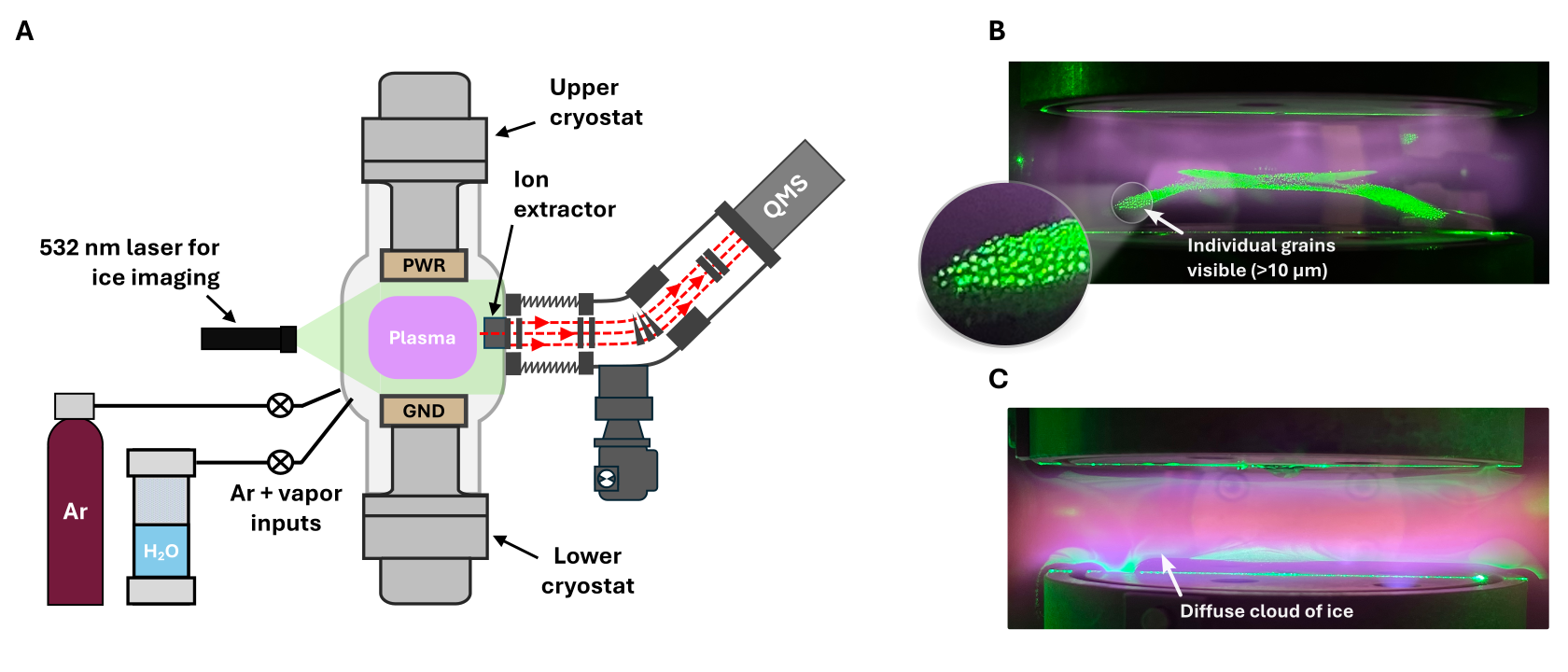}
    \caption{(A) Schematic of the experiment. (B) Image of laser light scattered off ice grains (green) suspended within the plasma (purple) at 130 K; individual grains are visible, and are enlarged in the overlay at left. (C) Image of scattered laser light at 170 K; the grains are much smaller and scatter the light diffusely.}
    \label{fig:schem-and-laser}
\end{figure}

A capacitively-coupled plasma device, shown schematically in Fig. \ref{fig:schem-and-laser}, produces a weakly-ionized plasma in a 40 Pa,  10:1 $\mathrm{Ar + H_2O}$ mixture. Argon is utilized as a chemically inert buffer gas, providing a thermal bath and a source of free electrons while isolating the nucleation pathway to water-derived species. The parallel-plate electrodes are each mounted on a cryostat and are independently set to temperatures between 50 K and 300 K. As the pressure is below the water triple point, ice nucleates directly from the gas phase at sufficiently cold temperatures.

When the electrodes are colder than 190 K, ice grains are observed via laser scattering in the plasma \citep{shimizu2010synthesis, chai2015formation}. The ice grains acquire charge through electron collection and are thereby confined within the bulk plasma; they exhibit complex motion coupled to ion and gas dynamics as their size and shape evolve via accretion or sublimation of water molecules and ice clusters \citep{chai2016vortex, marshall2017identification, nicolov2024phase, nicolov2025dynamics}. The ultimate grain size depends strongly on temperature: below 150 K, grains grow over several minutes to hundreds of microns in length, forming unique non-spherical morphologies \citep{nicolov2024phase, nicolov2025dynamics}, while at higher temperatures the grains remain much smaller, presumably as a result of an increased sublimation rate. 

A vertically-oriented sheet of 532 nm laser light passing through the   plasma illuminates  the ice grains confined in the plasma. Figure \ref{fig:schem-and-laser} B shows an image of both the plasma and the laser light  scattered from the ice grains at 130 K while Figure \ref{fig:schem-and-laser} C shows an image where the temperature is instead 170 K. At 130 K, the laser-illuminated ice grains are bright and easily distinguished from each other  indicating that they are much larger than the laser wavelength. At 170 K, the scattering is dimmer and more diffuse, indicating that the ice grains are comparable to or smaller than the laser wavelength. No laser scattering is observed at electrode temperatures above 190 K, indicating that for this temperature, the grains are $\lesssim 100$ nm which is  too small to scatter significant light. Nanometer-scale ice grains may exist above 190 K but would be invisible to the laser scattering diagnostic. The critical temperature at which  ice nucleation occurs thus lies somewhere above 190 K.

\subsection{Observation of ion clustering}
At the earliest stages of ion-induced nucleation, $\mathrm{H_2O}$ molecules cluster around plasma ions. To detect these precursors to ice grain formation, a quadrupole mass spectrometer (QMS) was situated at the plasma edge. Because the plasma has a positive potential, positive ions stream outward to the QMS and are easily detected, whereas negative ions are confined to the plasma interior and cannot be detected. The RF power sustaining the plasma was therefore modulated by a square wave at 500 Hz, and the QMS was synchronized to detect positive clusters (water molecules clustered around a positive ion) streaming from the powered plasma and negative clusters released during the afterglow \citep{howling1994time}. 

\begin{figure}
    \centering
    \includegraphics[width=\linewidth]{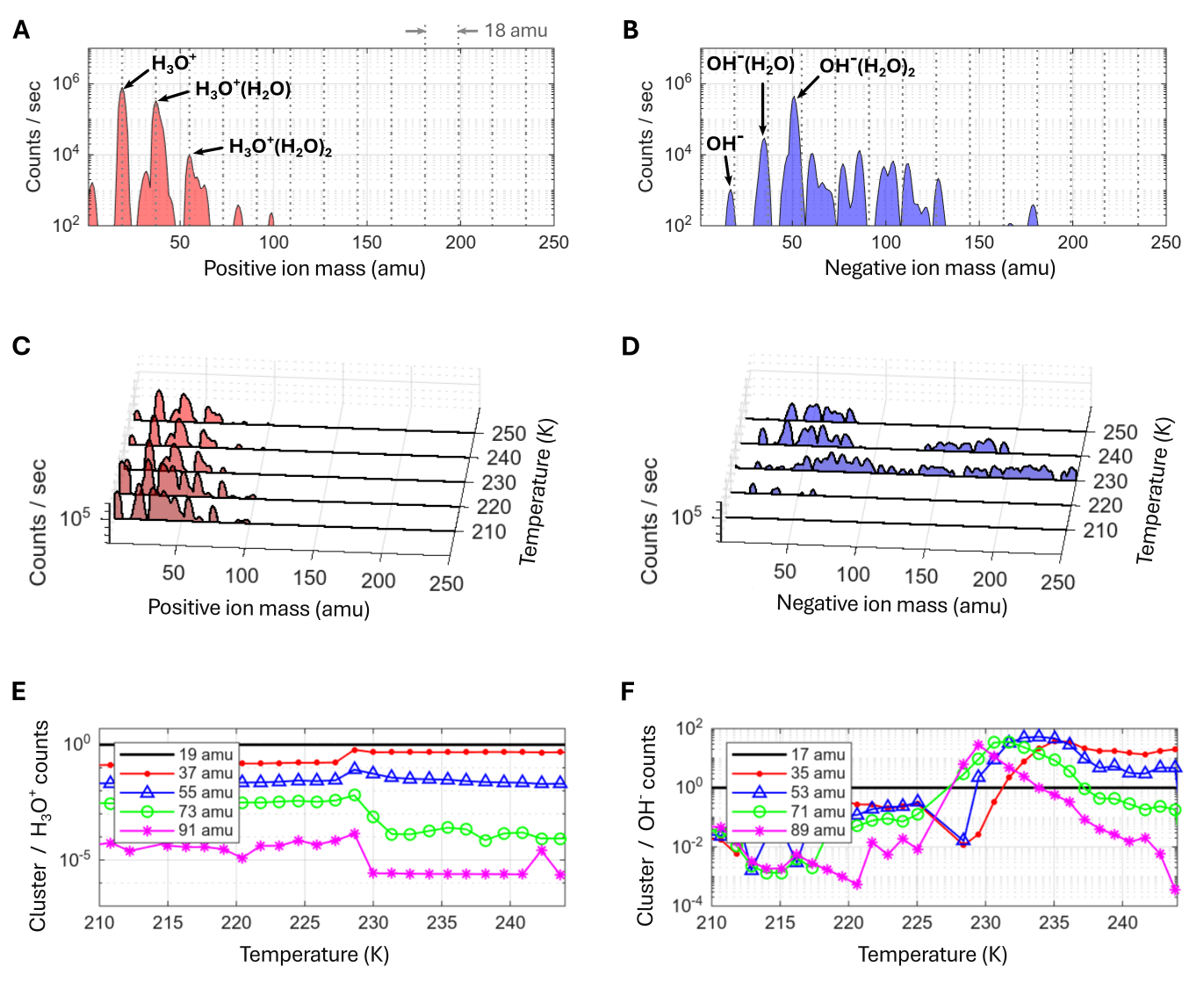}
    \caption{Mass spectra of positive clusters (A) and negative clusters (B), collected at 245 K, and smoothed with a 6-amu gaussian window to show long-range behavior. (C) and (D) show the variation of mass spectra with temperature, from 210 to 250 K. Concentrations of clusters about bare $\mathrm{H_3O^+}$ (E) and $\mathrm{OH^-}$ (F) with sizes $n = 1$ to 4, relative to concentration of the bare ion,  are plotted as temperature is steadily decreased.}
    \label{fig:mass-spec-clustering}
\end{figure}

Mass spectra collected at 245 K are plotted in Fig. \ref{fig:mass-spec-clustering} (A) and (B). The spectra are smoothed with a 6-amu Gaussian window to show large-scale structure of the spectrum, and intensities are corrected to account for the mass-sensitivity of the instrument using a  factor given in an application note for the instrument \citep{HidenAppNote282}. The spectra of both positive and negative clusters show a sequence of peaks with an approximately 18 amu  periodicity, indicating successive clustering of $\mathrm{H_2O}$ molecules around ions. For negative clusters, this extends to the upper mass limit of the plot (250 amu), indicating negative clusters containing over  $10$ attached water molecules. The mass of  positive clusters, meanwhile, only goes up to about 100 amu. 

The 18 amu periodicity is not perfect; the negative clusters do not strictly follow $\mathrm{OH}^{-}(\mathrm{H_2O})_n$. Above 100 amu, some of the largest peaks are separated by 17 or 19 amu rather than 18, suggesting that negative clusters of this size are more stable if they collect an H or OH radical rather than an $\mathrm{H_2O}$ molecule. This is likely due to the precise bonding geometry of the molecules in the cluster, related to known ``magic numbers" of $\mathrm{H_2O}$ at which clusters are significantly more stable than others  \citep{lee2005magic, bryantsev_Goddard2009evaluation}.

Although the density of the neutral argon atoms was significantly higher than that of water vapor, $\mathrm{Ar^+}$ did not constitute the majority of the positive ions at 245 K. Instead, ions originating from the decomposition of water dominate both positive and negative mass spectra. Specifically, $\mathrm{H_3^+}$, $\mathrm{O_2^+}$,   $\mathrm{H_3O^+}$,  $\mathrm{H^-}$, $\mathrm{OH^-}$, and $\mathrm{O_2^-}$ are prominently detected; of these, $\mathrm{H_3O^+}$ and $\mathrm{OH^-}$ tend to dominate, which is expected as these are produced at the highest rates for the electron energies in the experiment \citep{cottin1959etude}.

\subsection{Cluster stability with temperature}
In a separate experimental sequence, the plasma was maintained at constant gas and water flow while the temperature of the electrodes was slowly decreased from 250 to 210 K. Mass spectra were continuously collected as the temperature dropped to determine any changes in clustering behavior across the freezing point.  Figures \ref{fig:mass-spec-clustering} C and D show the positive and negative cluster mass spectra at 250, 240, 230, 220, and 210 K.  Panels E and F show the concentration of $\mathrm{OH^- (H_2O)_n}$ and $\mathrm{H_3O^+ (H_2O)_n}$ for $n = 1-4$, normalized to the concentration of the bare $\mathrm{OH^-}$ or $\mathrm{H_3O^+}$ ions.


Figure \ref{fig:mass-spec-clustering} C and E show that positive clusters do not change much when the temperature is lowered. While their concentrations change slightly  in the range of 225-235 K, with $\mathrm{H_3O}^+(\mathrm{H_2O})_3$ in particular becoming noticeably more abundant, positive clusters do not change relative to each other: the bare $\mathrm{H_3O^+}$ remains the most abundant at all temperatures, while each successive hydrate is less abundant than the last. This suggests that positive  clusters have little role in ice nucleation.

Figure \ref{fig:mass-spec-clustering} D and F show that negative  clusters have quite different behavior as the temperature dependence of negative clusters is highly nonlinear. Above 236 K,  $n=1$ (i.e., 35 amu) is the most stable cluster. As the temperature decreases from 236 to 225 K, larger clusters become stable, while smaller ones decrease in abundance. In particular, Figure \ref{fig:mass-spec-clustering}  F shows that at 233 K, the concentration of the $n = 2$ (i.e., 53 amu) negative cluster reaches a peak and dominates the other negative clusters, while $n = 1$ concentration drops sharply. At 230 K, $n = 3$ becomes the most abundant negative cluster, while $n = 2$ drops in concentration, and at 227 K, $n = 4$ dominates. Below 225 K, all the negative  clusters in the measurement range become significantly less abundant compared to bare hydroxide (OH$^-$). 

A striking feature of these measurements is that there is a 35 K gap separating the 225 K temperature at which the sudden depletion of negative ion clusters begins and the 190 K temperature at which optical detection  of ice grains begins. The existence of this 35 K gap separating these two  thresholds is  consistent with a nucleation-then-growth sequence where  clusters exceeding the  critical size associated with 225 K  rapidly accrete water molecules to suddenly achieve optically-detectable sizes. Thus, the threshold temperature for optical detection is  naturally  lower  than that for cluster depletion. Since the largest clusters that can be resolved by mass spectrometry have $ n \approx10^2$ molecules while  the smallest ice grains   detectable by laser scattering have a radius of the order of the laser wavelength corresponding to  $ n \approx 10^{10}$ molecules, the 35 K gap  corresponds to the region of extremely fast thermodynamically unstable growth  from  $10^2$ molecules to $10^{10}$ molecules, rather than a gap in the nucleation mechanism.

\section{Theoretical model of clustering}
\label{sec:Theoretical model}

Classical nucleation theory (CNT) predicts nucleation behavior by computing the change in Gibbs free energy $\Delta G$ for forming a spherical $\mathrm{H_2O}$ cluster around an ion using equilibrium thermodynamics and bulk properties of the stable condensed phase. Despite frequent criticism of its simplifying assumptions \citep{girshick2024nucleation, svensmark2007experimental, arnold1980ion}, CNT remains the most widespread method of predicting the stability of water clusters, their tendency to grow or shrink, and their production rate from vapor. 

In this framework, the Gibbs free energy to form a cluster of radius $r$ around a central OH$^-$ ion is given by:
\begin{equation}
    \Delta G(r) = -\left[\frac{4}{3} \pi r^3 \frac{\rho_\mathrm{ice}}{m_{H_2O}}   k_B T \;\ln\left(\frac{P_{H_2O, \infty}}{P_{sat}(T) } \right) + \frac{r \rho_{\mathrm{ice}}}{3 m_\mathrm{H_2O}} \frac{ |q_i| p}{\epsilon_0} \right] + \gamma 4 \pi r^2 - \frac{q_i^2}{8 \pi \epsilon_0} \left(1 - \frac{1}{\epsilon_r} \right) \left(\frac{1}{r_i} - \frac{1}{r} \right).
    \label{eq:Gibbs-eqn}
\end{equation}

Here, $\rho_\mathrm{ice}$ is the density of the condensed phase, $m_{H_2O}$ is the mass of a single water molecule,  $P_{H_2O, \infty}$ is the partial pressure of water far from the cluster, $P_{sat}$ is the saturated water vapor pressure, $\epsilon_r$ is the relative permittivity of the water cluster, $p$ is the dipole moment of a single water molecule,  $r_i$ is the radius of the OH$^-$ ion, and  $q_i$ is its charge. 

Equation \ref{eq:Gibbs-eqn} contains three contributions: a volume (chemical potential) term, which drives growth when the partial pressure exceeds the saturated vapor pressure; a surface energy proportional to the curvature-corrected surface tension  $\gamma$  \citep{kritikos2025can}; and an electrostatic term from the ion-dipole attraction \citep{girshick2024nucleation}. The volume term also includes a Boltzmann-like enhancement of water vapor density near the ion due to dipole attraction. This term which was not taken into account by \citet{gumbel2009charged} shows that the density of water vapor at the surface of the ion is greater than the water vapor density far away by a factor $\exp(pq_i/4 \pi \varepsilon_0 r^2 \kappa T )$ which  is very large, as shown by \citet{nielsen2011charge} to explain observation of nucleation in an ostensibly undersaturated cloud. Thus, the density of water molecules in the immediate vicinity of the OH$^-$  ion is orders of magnitude higher than at large distances from the ion; this means that the flux of water molecules towards the ion is much larger than would be the flux towards a same-size neutral particle.  

\begin{figure}
    \centering
    \includegraphics[width=\linewidth]{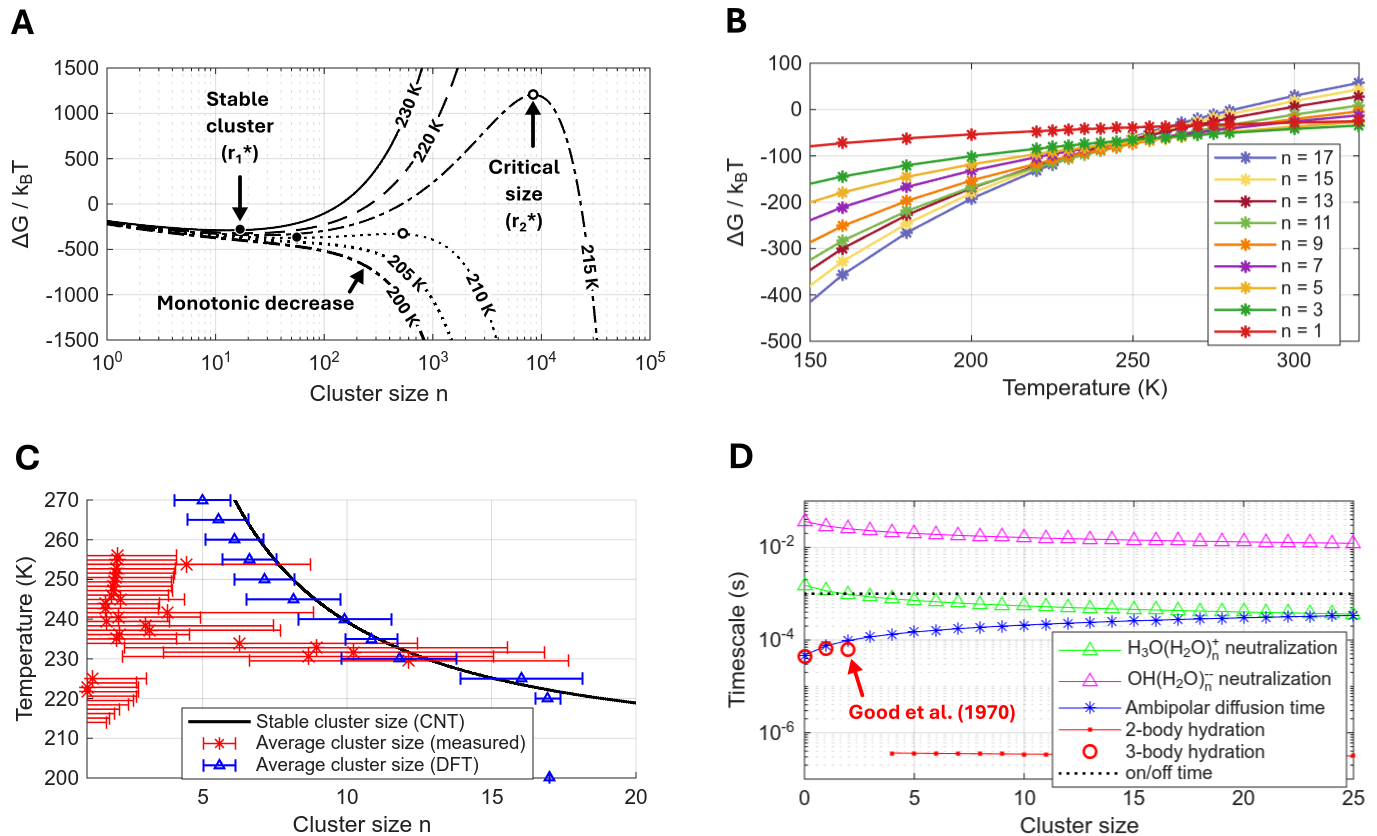}
    \caption{Comparison between classical theory, DFT simulations, and experimental results. A: the Gibbs free energy predicted by Eq.(\ref{eq:Gibbs-eqn}) using $P_{H_2O, \infty} =4$ Pa, $P_{sat}(T)$ given by \cite{murphykoop2005review}, $p=6.19 \times 10^{-30}$ C-m,  $\gamma= 0.109/(1+2\delta/r)$  N/m with $\delta=-0.4$ $\mathring{A}$, $r_i=1.1$ $\mathring{A} $, $\varepsilon_r \gg 1 $. B: The Gibbs free energies, computed from DFT, for several clusters. For clarity, only odd-numbered clusters are plotted, and only the lowest-energy configuration for each cluster size. C: The stable cluster size satisfying $d\Delta G/dn = 0$ from Eq. \ref{eq:Gibbs-eqn} (black), the average cluster size measured in experiment (red), and the average cluster size predicted by DFT (blue). Error bars on experiment and DFT sizes indicate the spread of the cluster size distribution. D: timescales associated with positive and negative ion neutralization, ambipolar diffusion of positive ions, and ion hydration, and the 1 ms on/off time of the RF power modulation. The 3-body hydration rates for $n = 0-2$ are taken from \cite{good1970ion}.}
    \label{fig:theory}
\end{figure}

Because Ostwald's step rule indicates that ice pre-nuclei are amorphous, we use the saturated vapor pressure over supercooled liquid rather than crystalline ice (see Eq.10 in \citet{murphykoop2005review}).  Figure \ref{fig:theory} A shows $\Delta G$ plotted for various temperatures, with cluster size expressed as the number of water molecules $n$ rather than radius  $r$, calculated assuming the cluster density is that of amorphous ice. The grain radius is related to its number of water molecules by $r= 2\times 10^{-10} n^{1/3}$ m so an ice grain with 1000 molecules would have a 2 nm radius.

The critical points of $\Delta G$ (where $d(\Delta G)/dr = 0$) indicate equilibrium cluster sizes. At high temperature, the vapor is undersaturated  ($p_{\mathrm{H_2O}} < p_\mathrm{sat}$) and the curve has a single minimum at an $n \sim 1$. As the temperature drops and the vapor becomes supersaturated, $\Delta G$ develops two critical points: a stable equilibrium $r_1^*$, corresponding to clusters of $n \sim 1-10$ water molecules, and a larger unstable equilibrium $r_2^*$, above which further growth is spontaneous. Ions will thus spontaneously hydrate to $r_1^*$; for further growth to occur, the cluster must climb an energy barrier, called the Kelvin barrier,  of $\Delta G(r_2^*) - \Delta G(r_1^*)>0 $.  As the temperature decreases, however, the two critical points converge, the Kelvin barrier shrinks, and at $T \approx 210$ K it vanishes. Below this temperature,  $\Delta G$ decreases monotonically with increasing $n$. Single ions then grow spontaneously into stable ice grains -- this spontaneous growth constitutes nucleation.

This process means that clusters undergo rapid, runaway growth below 210 K, and thus one would not expect to observe clusters with $n<10^3$ in any significant amount: all clusters would have rapidly fallen down the ``hill" in $\Delta G$, as seen in Fig. \ref{fig:theory} A. This explains the disappearance of the negative clusters in the experiment at $T < 220$ K; the stable clusters observed at higher temperature have all fallen down the hill in potential energy and grown beyond the mass range of the QMS.

\subsection{Comparison to heterogeneous nucleation}

To compare Eq.\ref{eq:Gibbs-eqn} to the classic no-plasma situation, we briefly summarize the discussion  given in \citet{gumbel2009charged}.
If $q_i$ is set to zero in Eq.\ref{eq:Gibbs-eqn}, the result is
\begin{equation}
    \Delta G(r) = -\frac{4}{3} \pi r^3 \frac{\rho_\mathrm{ice}}{m_{H_2O}}   k_B T \;\ln\left(\frac{P_{H_2O, \infty}}{P_{sat}(T) } \right) + \gamma 4 \pi r^2  .
    \label{eq:hetero}
\end{equation}
 At infinitesimal $r$, $d/dr \Delta G(r) $ is positive. As $r$ increases,  $d/dr \Delta G(r) $ decreases to zero at a critical radius $ r^*$, and then becomes negative for $r>r^*$. Thus at $r=r^*$,  $\Delta G(r)$ is at a  maximum and then, with increasing $r$,  declines for $r>r^*$. This maximum of $ \Delta G(r)$ constitutes the Kelvin barrier  that grains with $r< r^*$ cannot pass; after a while these too-small  particles evaporate and disappear. To have continuous growth to macroscopic size it is therefore necessary to start with particles having $r> r^*$ (continuous growth corresponds to falling off the energy cliff). This $r> r^*$ requirement for continuous growth corresponds to heterogeneous nucleation where  some pre-existing spherical substrate  with $r>r^*$ allows the water ice grain to grow continuously. In this situation a nucleation rate can be calculated having  the form $J=A \exp({-\Delta G(r)/ k T})$ which is a measure of the probability of there being sufficient energy to exceed the Kelvin barrier. Comparing this situation  to the finite $q_i$ situation where below a critical temperature the Kelvin barrier  disappears for all $r$, it is seen that the finite $q_i$ situation has an effective infinite nucleation rate as grains starting with arbitrarily small $r$ always experience a negative sloping $ \Delta G(r)$ with magnitude that  starts off  small  but then becomes large. This means that for negative ions below a critical temperature there is no Kelvin barrier to overcome for all clusters, no matter how small.

\subsection{Cluster stability from DFT calculations}
CNT treats clusters as a continuous medium and omits quantum effects, assumptions that break down for clusters of only 1–10 molecules. To address this, we performed Density Functional Theory (DFT) computations, a quantum mechanical method that computes atomic-scale energetics from first principles, to obtain the Gibbs free energy of $\mathrm{OH^{-}(H_2O)_n}$ clusters for  $n=1 – 17$ at experimental temperatures. The free energy to form a cluster from vapor is then calculated as:
\begin{equation}
    \Delta G = G_{{OH^-(H_2O)_n}} - G_{{OH^-}}-n G_{{H_2O}},
    \label{eq:dG-DFT}
\end{equation} 
the energy difference between the cluster and its separate component parts.

Initial cluster geometries were taken from Refs.~\cite{Zanuttini2015,Lemke2021} and optimized at the B3LYP-D3/6-311++G** level using ORCA \cite{ORCA}, with thermochemistry computed following Refs. \cite{Grimme2012,Herzberg1945}. Results were corrected from atmospheric pressure to the experimental water partial pressure of 4 Pa via:
\begin{equation}
    \Delta G(p)=\Delta G(p^0) - n k_B T \ln\frac{p}{p^0},
    \label{eq:pressure-correction}
\end{equation}

\noindent where $p^0$ is the atmospheric pressure and $p$ is the 4 Pa partial pressure of water in the experiment. This correction applies the standard ideal gas chemical potential shift; we note it does not account for non-uniformities in the local water vapor density near the ion due to dipole attraction, or depletion due to rapid ice deposition.

Figure~\ref{fig:theory}B shows $\Delta G$ versus temperature for the minimum-energy configuration of each cluster. Larger clusters decrease more steeply with temperature, so the most stable cluster size grows rapidly with cooling:  $n=3$  is favored at 300 K,  $n=5$  at 250 K, and  $n=17$  (the largest simulated) at 220 K. The probability of forming a cluster of size $n$ is proportional to $\exp(-\Delta G_n/k_B T)$; this exponential thus gives a cluster size distribution that can be compared with experimental results.

\subsection{Comparison of predicted and measured clusters}
The critical cluster size from CNT is plotted  in Fig. \ref{fig:theory} C alongside the mean and the spread of the cluster size distributions obtained from DFT. While the shape and spread of the size distribution predicted by DFT is impacted by the details of the bonding geometry, the mean values follow the trend predicted by CNT with good agreement below 250 K.

These predictions are also compared to the measured mass spectra of negative clusters. The mass spectra were binned in 18 amu intervals corresponding to successive hydration states to obtain a cluster size distribution at each temperature. The mean and standard deviation of the size distribution are plotted in red in Fig. \ref{fig:theory} (C). As the temperature drops to a critical value around 225 K, the mean cluster size increases to $n \sim 15$. Below this critical temperature, the measured cluster size suddenly drops to near-zero as only $\mathrm{OH^-}$ is observed. This temperature dependence follows CNT predictions: larger clusters stabilize as temperature drops, until growth becomes spontaneous and clusters grow into macroscopic ice grains beyond the mass spectrometer's detection range.

While the trend of measured cluster size with temperature qualitatively matches the theory, the trend deviates quantitatively. Most notably, the measured mean cluster remains smaller than either CNT or DFT predict at higher temperatures; then, as the critical temperature is approached, the clusters grow more rapidly than predicted. The measured cluster size distributions are also broader than those predicted by DFT.

Both the deviations in the temperature trend and in the breadth of cluster sizes are likely due to the non-equilibrium nature of the plasma. As the DFT size distributions are computed from a Boltzmann factor, they implicitly assume an equilibrium forward and reverse molecular exchange between the cluster and its surroundings. In reality, cluster formation, growth, and destruction rates compete with ion hydration which populates non-equilibrium cluster sizes and broadens the distribution. At higher temperatures particularly, shallow energy differences between hydration states mean that evaporation and clustering rates are nearly balanced, and clusters may not reach equilibrium sizes before escaping the plasma. 

\subsection{Charge asymmetry of clustering kinetics}
Clusters of water molecules bound to positive ions in the experiment do not show the same temperature dependence as  clusters bound to negative ions: the most-abundant positive  clusters remain small, and the variation of concentrations does not correlate with ice grain formation. This suggests that positive clusters are not the progenitors of the observed ice particles.

Water aerosols have long been known to nucleate preferentially on negative ions \citep{wilson1897}. This is missed entirely by CNT, as Eq. \ref{eq:Gibbs-eqn} is agnostic to the sign of the charge, and while DFT does reflect the sign preference, the predicted rate differences are slight and heavily species-dependent \citep{nadykto2006quantum, svensmark2007experimental}-- neither model can account for the extreme charge sign preference observed in experiment. Instead, this preference is primarily a consequence of differing kinetic loss rates of positive and negative ion clusters in the plasma  \citep{bellan2022}; these rates do not allow sufficient time for thermodynamic equilibrium to be reached. The evolution of a cluster in the experiment is effectively a competition between \textit{hydration} through molecular accretion and \textit{destruction}, either through neutralization by opposite-sign charges or transport out of the plasma. Figure \ref{fig:theory} D shows estimates of the hydration and destruction rates for clusters of sizes $n = 1 - 25$. 

If a positively-charged cluster is neutralized by a free electron, or a negatively-charged cluster is neutralized by a positive ion, the cluster loses its stabilizing charge-dipole interaction and quickly evaporates. The neutralization via collision of an electron with a positive ion water cluster is known to be rapid \citep{ojekull2008dissociative}. The neutralization rate of a positive cluster by an electron depends on $\sqrt{T_e/m_e}$, and for a negative cluster by a positive ion, by $\sqrt{T_i/m_i}$. Since $T_e \gg T_i$ and $m_i \gg m_e$, positive clusters are thus neutralized thousands of times faster than negative clusters.

Furthermore, positive clusters are accelerated toward the electrodes by the ambipolar electric field, further shortening their lifetime. This field exists because plasma electrons escape much faster than plasma ions, so the plasma is positively charged and thus tends to expel positively charged particles while trapping negatively charged particles.  This dominates removal for the smallest positively-charged clusters due to their higher mobility.

While the binary collision rate between an ion cluster and a water molecule is very fast compared to the destruction rate, hydration of small clusters requires a third body (most commonly a neutral argon atom) to carry away excess binding energy and stabilize the cluster \citep{good1970ion}. While precise hydration rate constants for $\mathrm{H_3O^+(H_2O)_n}$ in argon buffer gas are absent from the literature, hydration rate constants measured in nitrogen for $n = 1-3$ \citep{good1970ion}, plotted in Fig. \ref{fig:theory} D, take about 2 orders of magnitude longer than the timescale for binary collisions, and are comparable to the timescale for ambipolar diffusion of positive ions in our apparatus. The requirement for a third body will continue until $n$ is sufficiently large that the vibrational modes can absorb the binding energy internally.

These effects together act as a kinetic filter: positive ion clusters escape the plasma via ambipolar diffusion faster than they can grow through three-body hydration, producing the sharp cutoff at  $n \approx 3$  observed in the mass spectrum. Negative ion clusters, by contrast, are electrostatically confined and neutralized only by the much slower ion-ion process. Their lifetime therefore greatly exceeds the hydration timescale at every stage of growth, allowing them to grow continuously into macroscopic ice grains.

\section{Implications and scaling to natural environments}
While ion clustering has long been postulated as a precursor to ice formation in cold, low-pressure environments such as Earth's mesosphere, various planetary environments, interstellar clouds, and protoplanetary disks \citep{arnold1980ion, bellan2022}, prior laboratory measurements have operated far from natural conditions \citep{ojekull2008dissociative, froyd2003experimental, good1970ion}. The present experiment more closely approximates these environments: the system pressure is below the water triple point and the temperature spans the range of rapid nucleation onset. Some differences remain-- notably, the pressure and water fraction are both elevated relative to most space environments-- but these affect rates rather than the underlying chemistry. The inert argon buffer gas influences diffusion and ionization but not the nucleation pathway itself; the higher water fraction accelerates clustering, and the small system size shortens the lifetime of positive clusters relative to natural settings.

The relatively high pressure argon buffer gas, which clearly differs from astrophysical or mesospheric conditions, was necessary for three reasons. First, this pressure was required for producing a steady-state plasma discharge without excessive RF power input that would heat the cryogenic gas. Secondly, the high pressure ensures that temperature remains low throughout the discharge, as heat can be rapidly conducted to the cryogenic electrodes. Finally, the high pressure is necessary for increasing diffusion times, so that clusters do not escape to the walls before they grow to appreciable size. As natural environments do not have nearby walls, no such pressure requirement exists, and particle formation continues at lower pressures albeit at lower rates.

Because argon is mainly inert, it has little effect on the nucleation mechanism, serving only as a thermal bath and a source of free electrons. While the total pressure affects the diffusion rate and thus  the cluster growth rate, it has no effect on the nucleation mechanism or on cluster stability. The cluster stability depends only on the ratio of the water vapor partial pressure to the saturated vapor pressure, such that clustering occurs when the vapor is supersaturated (see Eq. \ref{eq:Gibbs-eqn}). The partial pressure of water thus can be scaled to astrophysical conditions, affecting only the rates of the physical processes and not the pathways themselves, as long as supersaturation is preserved. Because saturated vapor pressure decreases dramatically with temperature \citep{murphykoop2005review}, in-space conditions are typically supersaturated even when the absolute value of the water vapor partial pressure is extremely low.

For ion-induced nucleation to proceed, the lifetime of charged clusters must be sufficiently  long  for the cluster to grow to a stable size. In the experiment, positive clusters are rapidly expelled to the walls by an ambipolar electric field, while negative clusters are perfectly confined so their lifetime is governed solely by neutralization caused by collisions with positive ions. No such charge-sign filtering by confinement properties exists in natural environments, and thus positive clusters can grow larger. However, because both positive and negative clusters are destroyed through neutralization by opposite-charged particles,  negative ions will still have significantly longer lifetimes than positive ions  because electron-positive cluster collision rates will be higher than ion-negative cluster collision rates, owing to the small mass and high velocities of electrons. Negative clustering thus retains its advantage in nucleation efficiency over positive clustering \citep{gumbel2009charged, bellan2022}. 

The competition between hydration and destruction is set to first order by the density ratio of water to ions,  $n_\mathrm{H_2O} / n_i$. When this ratio is large, ion hydration time ($\tau_{hyd.}$) is fast relative to neutralization ($\tau_\pm$) and clusters can grow; when $n_\mathrm{H_2O} / n_i$ is small, $\tau_{\pm} < \tau_{hyd}$, and ions are neutralized before accumulating enough water molecules to reach appreciable size. This same density ratio governs the strength of the charge sign preference: the sign preference remains prominent as long as $\tau_{hyd.}$ is small compared to $\tau_-$, while \emph{not} being negligibly small compared to $\tau_+$. As $\tau_+$ is about $10^2$ times smaller than $\tau_-$ due to the electron-to-ion mass ratio, the sign preference is notable over several orders of magnitude of $n_\mathrm{H_2O} / n_i$. 

In the experiment, $n_{H_2O} \approx 10^{15}$ cm$^{-3}$, while ion density is about $10^{9}$ cm$^{-3}$. Thus, $n_\mathrm{H_2O} / n_i \approx 10^6$. If $n_\mathrm{H_2O} / n_i << 10^6$, negative clusters are likely neutralized too quickly to induce nucleation, while if $n_\mathrm{H_2O} / n_i >> 10^6$, positive and negative ions can both seed nucleation.

\subsection{In Protoplanetary Disks}
In protoplanetary disks, most of the water beyond the snow line is frozen into icy grains and pebbles which settle
gravitationally to the disk midplane \citep{banzatti2023jwst, du2014water}. Observations have shown that water vapor abundance throughout the disk is lower than initial models expected, indicating ongoing ice formation across the disk's vertical layers \citep{du2017survey, laibe2014growing, bergin2010sensitive}. The ice is generally assumed to form heterogeneously around pre-existing
silicate or carbonaceous dusts mixed with the gas \citep{du2014water}.

Water vapor concentration and fractional ionization are both very low on the disk midplane \citep{ivlev2016ionization, potapov2021dust}. This is because water vapor rapidly deposits on larger surfaces while dust blocks incoming ionizing radiation. For a typical T-Tauri protoplanetary disk just beyond the snow line ($\geq1.5$ AU), fractional ionization is approximately $10^{-12}$ on the midplane, while water vapor concentration is estimated as $10^2-10^4$ cm$^{-3}$ \citep{du2014water, bergin2010sensitive}. These low concentrations overall would make ongoing ion-induced nucleation very slow compared to deposition on pre-existing particles, and as $n_{H_2O}/n_i \approx 1-100$, the remaining water would not be sufficient to form stable ion clusters before cluster neutralization. However, this situation changes in the outer layers away from the midplane, where ionization rates climb as radiation from the central star is less attenuated by  dust, and the abundance of water vapor and other volatiles increases in concentration due to desorption by FUV photons \citep{ivlev2016ionization, bergin2010sensitive}. Thus, ion-induced nucleation rates will rise. As dust particulates in outer layers are also smaller and fewer in number \citep{ivlev2016ionization}, heterogeneous nucleation rates decrease.

\begin{figure}
    \centering
    \includegraphics[width=\linewidth]{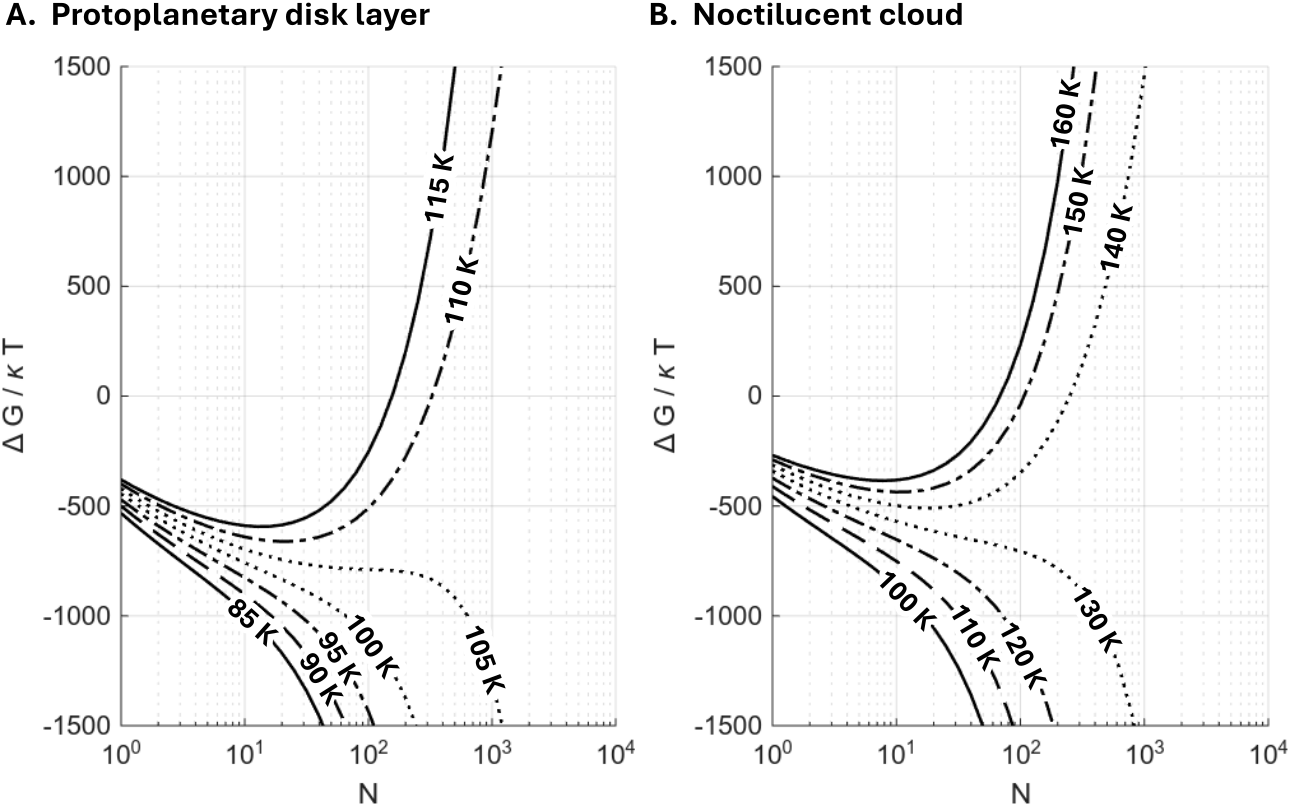}
    \caption{A: Gibbs free energy for nucleation about an OH$^-$ ion (Eq. \ref{eq:Gibbs-eqn}) at protoplanetary disk parameters corresponding to $R=20$ AU and $Z/R=0.2$, using values from \citet{agundez2018chemistry}, where the H density is $10^9$ cm$^{-3}$ and the relative water abundance is  $10^{-9}$. At this location, \citet{agundez2018chemistry} indicates $T = 70$ K, at which the plot is monotonically decreasing, and so ice will spontaneously nucleate on an OH$^-$ ion. B: Same as panel A, except for noctilucent cloud parameters at 85 km, where air pressure is 0.35 Pa and water vapor is 2 parts per million, as given by \citet{gumbel2009charged}}
    \label{fig:dG_disks_clouds}
\end{figure}

For a quantitative example, we consider the nominal parameters from \citet{agundez2018chemistry}  on a T-Tauri disk  at  $R=20$ AU and $Z/R=0.2$.  At this location Fig. 2 of \citet{agundez2018chemistry} indicates a 70 K  gas temperature and a $10^9$ cm$^{-3}$ gas number density, while Fig. 4 of \citet{agundez2018chemistry} gives the fractional water abundance to be $10^{-9}$. The H$_2$O number density is thus $1$ cm$^{-3}$ and so $P_{H_2O, \infty} = 9.7 \times 10^{-16}$ Pa. At 70 K, the Murphy-Koop expression for saturated vapor pressure gives $P_{sat}=  9 \times 10^{-25}$ Pa  so $P_{H_2O, \infty}/P_{sat} = 10^9$ \citep{murphykoop2005review}; thus even though  $P_{H_2O, \infty}$ is very low, $P_{sat}$ is even lower so the situation is supersaturated. Figure \ref{fig:dG_disks_clouds}A plots Eq.\ref{eq:Gibbs-eqn} for a range of temperatures and it is seen that the Kelvin barrier does not exist for temperatures below about 85 K. Thus, the 70 K temperature at $R=20$ AU and $Z/R=0.2$ satisfies the condition for monotonic decrease of $\Delta G$, and hence nucleation on  OH$^-$ negative ions will occur spontaneously.

This negative ion-induced nucleation process competes with nucleation on silicate dust grains. The relative importance of each nucleation pathway depends on the ratio of silicate dust grains to OH$^-$ negative ions, and on the effective cross-section of each potential nucleation seed (ion or dust). Although the total surface area of OH$^-$ ions in a given volume may be much smaller than the total surface of uncharged silicate grains in the same volume, the flux of water molecules on an OH$^-$ ion will be amplified because water molecules, having a large permanent dipole moment, are attracted to negative ions. 

To estimate the relative importance of heterogeneous nucleation and negative ion-induced nucleation by OH$^-$, we again use T-Tauri protoplanetary disk parameters from \citet{agundez2018chemistry}. We assume the dust is silicate, a background gas of molecular hydrogen, and a dust-to-gas mass ratio of 10$^{-2}$. We further assume that the dust radii follow a $r^{-3.5}$ MRN size distribution for 5 nm $<r<$ 1000 nm \citep{mathis1977size}, and that relative abundances of silicate and water are the same. Because the minimum silicate grain size of 5 nm exceeds the Kelvin barrier, all silicate grains can act as nucleation centers. 

Two factors to be discussed below significantly increase the importance of negative ion nucleation compared to nucleation on silicate grains, namely: (i)\ reduced effective surface area of silicate grains compared to the total surface of isolated  individual silicate molecules having the same total mass as a grain, and (ii)\  increased effective surface area of negative ions through electric lensing.

\subsubsection{Effective area of silicate grains}
Unlike OH$^-$, which is a nucleation site composed of a single free molecule, silicate grains are compact particles with only a small fraction of their molecules exposed at the surface. Because the ratio of surface area to volume decreases with $N^{-1/3}$, where $N$ is the number of molecules in the nucleation seed, the total surface area available for nucleation per unit mass (or molecule) is significantly smaller for heterogeneous nucleation on dust compared to that of ion-induced nucleation.

To calculate the reduction in surface area compared to individual molecules, we apply the  $r^{-3.5}$ MRN size distribution \citep{mathis1977size}. The MRN distribution has  $r_{1}=5 $ nm  and $r_{2} =1000 $ nm respectively as the smallest and  largest grain sizes in the distribution so the probability of a grain having a radius between $r$ and $r+dr$ is 
\begin{equation}
P(r)dr=\frac{r^{-3.5}dr}{\int_{r_{1}}^{r_{2}}r^{-3.5}dr}.
\label{probability}
\end{equation}
The effective surface area of the grains capable of nucleating is%
\begin{equation}
\left\langle S_{nuc}\right\rangle =\int_{r_{1}}^{r_{2}}4\pi r^{2}P(r)dr
\label{avg sfc}
\end{equation}%
while the average number of molecules in a grain is%
\begin{equation}
\left\langle N\right\rangle =\frac{\int_{r_{1}}^{r_{2}}\frac{4\pi r^{3}}{3}P(r)dr%
}{\frac{4\pi r_{\mathrm{Si}}^{3}}{3}}.  \label{avg N}
\end{equation}%
Let $\sigma _{\mathrm{Si}}=4\pi r_{\mathrm{Si}}^{2}$ be the cross section of a
single silicate molecule. If the average grain were decomposed into isolated
individual molecules, the surface area of all these molecules would be 
\begin{equation}
S_{decomposed}=\left\langle N\right\rangle \sigma _{\mathrm{Si}}=\frac{%
\int_{r_{1}}^{r_{2}}4\pi r^{3}P(r)dr}{r_{\mathrm{Si}}}.  \label{decomposed}
\end{equation}%
The ratio $\alpha $ of the effective surface area to the surface area of the
same grains in a decomposed state of isolated individual molecules is 
\begin{equation}
\alpha =\frac{\left\langle S_{nuc}\right\rangle }{S_{decomposed}}=r_{\mathrm{Si%
}}\frac{\int_{r_{1}}^{r_{2}}r^{2}P(r)dr}{\int_{r_{1}}^{r_{2}}r^{3}P(r)dr}.
\label{alpha}
\end{equation}%

\bigskip Evaluation of the integrals in Eq.\ref{alpha} gives 
\begin{equation}
\alpha =\frac{r_{\mathrm{Si}}\left( r_{1}^{-1/2}-r_{2}^{-1/2}\right) }{%
r_{2}^{1/2} -r_{1}^{1/2} }\simeq \frac{r_{\mathrm{Si}}}{\sqrt{r_{1}r_{2}}}.
\label{silicate reduction}
\end{equation}%
Using  $r_{\mathrm{Si}}=0.25$ nm gives $%
\alpha =\allowbreak 3.5\times 10^{-3}$ so the effective surface area for nucleation of an MRN\ distribution of silicate grains is about 300 times smaller than the surface area obtained if the grains were decomposed into individual isolated molecules.

\subsubsection{Charge-enhanced effective surface area of hydroxide negative ions}
Water is dissociated by UV radiation at 200 nm to produce OH, and thus the ratio of OH to water depends on height above the midplane. Figure 4 of \citet{agundez2018chemistry} shows that at  $R=20$ AU \ and $Z/R=0.2$, the OH/H$_{2}$O ratio is $10^{-2}$.  The dominantly hydrogen gas is continuously ionized to produce free electrons, and these electrons come to thermal equilibrium with the 70 K gas \citep{ivlev2016ionization}. Since the electron temperature of 70 K is much less than the OH electron affinity of 1.8 eV \citep{bellan2022}, the OH population is nearly completely converted into OH$^{-}$ over time, such that the OH$^-$/H$_{2}$O ratio is also $10^{-2}$. 

Because water molecules are attracted to hydroxide negative ions, the effective capture cross section of each OH$^-$ ion exceeds its geometric cross section $\pi r_{i}^{2}$. We estimate this enhancement by generalizing the orbital-motion-limited method \citep{allen1992probe} to a dipole interaction: an incoming water molecule at distance $r$ from the ion has potential energy $-\left\vert pE(r)\right\vert$, where $p$ is the water dipole moment and $E(r)=q_i /4 \pi \varepsilon_0 {r}^2$ is the electric field produced by the negative ion. Defining $v_\infty$ as the nominal velocity of a water molecule when it is  far from the negative ion, energy conservation gives:
\begin{equation}
\frac{1}{2}mv_{\infty }^{2}=\frac{1}{2}mv_{gr}^{2}-\left\vert
pE(r_{i})\right\vert,   
\label{OML energy}
\end{equation}
where $v_{gr}$ is the water molecule velocity when it makes a grazing impact on the negative ion. Let $b$ be the impact parameter of an  incoming molecule making a grazing collision. As in \citet{allen1992probe}, conservation of angular momentum gives $bv_{\infty }=r_{i}v_{gr}$, so Eq. \ref{OML energy} becomes:
\begin{equation}
\frac{1}{2}mv_{\infty }^{2}=\frac{1}{2}mv_{\infty }^{2}\frac{b^{2}}{r_{i}^{2}}-\left\vert pE(r_{OH})\right\vert,   \label{OML}
\end{equation}%
which gives:%
\begin{equation}
\frac{\pi b^{2}}{\pi r_{i}^{2}}=1+\frac{\left\vert pE(r_{OH})\right\vert }{\kappa T}%
\simeq 760.  \label{OH enhancement}
\end{equation}

Combining the 300-fold reduction of effective surface of silicate grains given by Eq. \ref{silicate reduction} with the 760-fold enhancement of effective hydroxide cross section given by Eq.\ref{OH enhancement} shows that if the hydroxide and silicate abundances were the same, the effective cross section of hydroxide would be greater by a factor of $2 \times 10^{5}$.  Figure 4 of \citet{agundez2018chemistry} indicates that the actual ratio of hydroxide to water is about $10^{-2}$ at R= 20 AU, Z/R =0.2, so assuming equal abundance of silicate molecules and total water molecules, nucleation via hydroxide negative ions is 2000 times more important than nucleation via silicate grains. 

We further note that this estimate does not account for vertical settling of larger silicates within the disk, which may cause a lower local dust admixture than the constant dust-to-gas mass ratio assumed \citep{laibe2014growing}, further increasing the relative importance of negative ion-induced nucleation. Future modeling of disks would undoubtedly take into account more subtle effects than discussed here and so better constrain the relative rates of hydroxide-induced and silicate-induced ice growth.

\subsection{In the Earth's polar mesosphere}
In the summer polar mesosphere at 85 km altitude the temperature is about 130 K and water vapor nucleates to form small ice grains that constitute noctilucent clouds. The air pressure is 0.35 Pa and the  water concentration is 2 ppm  \citep{gumbel2009charged} so   $P_{H_2O, \infty}=7\times 10^{-7}$ Pa. In addition,  $(n_\mathrm{H_2O} / n_i)_{meso.} \approx 10^{4}-10^{6}$ so ion hydration is fast compared to ion neutralization.  The nucleation kinetics are thus similar to  the Caltech experiment, indicating  efficient clustering and a sign preference for negative ions over positive ions. Figure \ref{fig:dG_disks_clouds} B plots Eq.\ref{eq:Gibbs-eqn} for noctilucent cloud parameters and shows that nucleation of water molecule clusters on OH$^-$ anions has a threshold temperature around 130 K (i.e., where $\Delta  G$ is monotonically descreasing).  This temperature is thus consistent with the observed  temperature at which noctilucent clouds form.

Both heterogeneous and ion-induced nucleation mechanisms have been proposed to explain this mesospheric ice. \citet{gumbel2009charged}  showed that positive-ion-induced nucleation cannot explain the observed ice nucleation because of quick cluster destruction by free electrons, while recent rocket-borne measurements have shown a truncated mass spectrum of positive particulates compared to negative particulates \citep{stude2025measurement}. These are both in agreement with the findings in the Caltech experiment.

Heterogeneous nucleation, meanwhile, relies on meteoric smoke particles (MSPs),  hypothesized to be nanometer-scale particulates at the relevant altitude that are the seed for the  10-50 nm size mesospheric ice grains \citep{havnes2014size, plane2018impacts}. \citet{gumbel2009charged} showed that neutral MSPs are too small to overcome the Kelvin barrier, and thus only charged MSPs could efficiently seed nucleation at the measured rates. As the charge and particulate size vary substantially through the mesosphere with altitude and time of day \citep{robertson2014detection}, it is unclear whether the charged MSP fraction can explain the ice nucleation rate. 

Negative ions have been mostly overlooked for nucleation as previous charged particle investigations focused on the heavy negatively-charged MSPs. However, negative ions derived through water and air chemistry, especially $\mathrm{O_2^-}$, $\mathrm{O^-}$, $\mathrm{NO_3^-}$, and $\mathrm{CO_3^-}$ should be prevalent \citep{kazil2003university}. Furthermore, molecules ablated from MSPs  are also highly electronegative and so could contribute to the negative ion population \citep{guiu2025oxygen}. Rocket-borne measurements have shown populations of negatively-charged particles in the $m/Z$ < 400 amu range, having number densities that are orders of magnitude greater than that of  MSPs at 80 km, reaching nearly $10^5$ cm$^{-3}$ \citep{robertson2014detection, stude2025measurement}. These light particulates could be explained by water clustering around light negative ions such as the abundant O$_2^-$, though poor mass resolution of the measurements below 100 amu has prevented definitive identification of small ion species. These measurements are often assumed to indicate remnants of MSP disintegration rather than signals of ongoing ion-induced ice nucleation. Whether these light negative particles are produced ``homogeneously" through $\mathrm{H_2O}$ dissociation, through chemistry of other atmospheric gases, or through ablation of micrometeorites, their high abundance and identical charge compared to nm-scale charged meteoric smoke suggest that they should be the dominant pathway for ice formation.

\section{Summary and conclusions} 
We have experimentally demonstrated spontaneous water-ice nucleation in a weakly ionized plasma via successive hydration of negative ions. This is the first  observation of simultaneous molecular precursors and macroscopic ice grain formation in a single controlled system and thus links microscopic to macroscopic scales that have previously only been studied in isolation. Our results and analysis demonstrate that ion-induced nucleation is an efficient path to ice grain formation for environments where cold, weakly ionized gas and water vapor coexist. This path requires no particulate seeds, no stable liquid phase, and no exotic chemistry. 

The agreement between the measured critical temperature, CNT predictions, and DFT simulations indicates that clusters of water molecules on negative ions undergo runaway growth near the temperature predicted by classical nucleation theory. The strong sign preference for negative  over positive ions demonstrates that plasma kinetics are an important modulator: positive clusters are fragmented by electron impacts before reaching appreciable size, whereas negative clusters are not subject to the same rapid destruction and so have the sizes and onset temperature that thermodynamics predicts. 

The experiments presented here use inert argon as a buffer gas, and thus restrict the water-clustering path to products of water molecules only. Natural environments may have additional reactive species such as molecular or atomic H, N and O that may alter hydration rates or cluster stability. Extending these experiments to mixed-gas environments would directly test how having different species modifies the efficiency of negative-ion-induced nucleation, opening a route to quantitative predictions for specific atmospheric and astrophysical settings.

Because the required conditions of cold background gas, modest water abundance, and weak ionization are ubiquitous across nature, spanning planetary atmospheres and protoplanetary disks,  nucleation mediated by negative ions can provide a route to water-ice formation in a variety of environments in the absence of foreign seed particles. Negative-ion nucleation occurs at lower electron energy and has fewer limitations than positive ion-induced nucleation due to the lower fragmentation probability of negative clusters. These results indicate that the abundance and reactions of negative ions should be considered in ice formation models of protoplanetary disks and planetary atmospheres. A quantitative estimate shows that nucleation induced by negative ions is plausible at the surface of a T-Tauri protoplanetary disk.

An important argument in favor of nucleation mediated by negative ions compared to heterogeneous nucleation is that  nucleation via negative ions does not require foreign seed particles and so requires fewer assumptions. Thus,  in the spirit of Occam's razor principle that the simplest explanation should be preferred,  nucleation via negative ions should be given serious consideration.

\section*{Acknowledgments} 
This work was supported by the National Science Foundation through NSF grant $\#$2308558. WAG additionally thanks the NSF (CBET 2311117) for funding.

\newpage
\appendix

\section{Methodological details}
\subsection{Plasma device specifications}
The plasma device is a capacitively-coupled discharge (CCP) in a 3.8-liter stainless steel vacuum chamber with 8-inch inner diameter. A turbomolecular pump backed by a diaphragm pump evacuates the chamber to a base pressure of $\sim10^{-6}$ Torr. Parallel-plate copper electrodes, separated by an adjustable gap, are driven at 13.56 MHz by an RF generator (T\&C Power Conversion AG 0113) through an automatic impedance matcher. In the experiments presented here, the lower electrode is powered to 30 W, and the upper is grounded.

Each electrode is mounted to the cold head of a single-stage Gifford-McMahon cryocooler (SHI Cryogenics CH-104). The electrode-cryocooler interface is a disk of sapphire, which electrically insulates the electrode from the cooler while providing good thermal conductivity. A resistive heater cartridge and silicon diode thermometer embedded in each electrode are managed by a PID temperature controller (Stanford Research Systems CTC100), which maintains independently-set electrode temperatures between 50 and 300 K. 

Argon is supplied from a standard cylinder through a precision leak valve at 50--100 SCCM. Water vapor is sourced from a sealed canister partially filled with distilled water from which air has been evacuated, leaving pure vapor at the saturated vapor pressure of $\sim$10 Torr in the space above the liquid. This vapor is metered into the chamber through a vapor mass flow controller (MKS Type 1150C) at 10 SCCM. Chamber pressure is set by balancing the gas inflow against the turbomolecular pump outflow via a partially-closed angle valve, and is monitored by a capacitance manometer (MKS).

\subsection{Laser scattering diagnostic}
Ice grain formation was monitored via laser sheet scattering. A 532 nm Nd:YVO$_4$ laser pointer was directed through a cylindrical lens to produce a vertical sheet. The sheet was directed through the center chord of the plasma device, illuminating a cross-section of the ice cloud suspended within the discharge. 

Scattered light was observed directly through a viewport oriented $90^\mathrm{o}$ to the laser sheet, and recorded photographically using a Nikon DSLR camera fitted with a macro wide-angle lens \citep{nicolov2024phase}. A 532 nm bandpass filter was used to confirm that observed scattering originated from the laser rather than plasma emission; the images presented in Fig. 2B were taken without the filter to show both the plasma and the ice simultaneously. 

Ice grains diffusely scatter some detectable light once their diameter exceeds roughly 20\% of the laser wavelength, corresponding to a detection threshold of $\sim$100 nm. The onset temperature of 190 K reported here reflects the temperature below which laser scattering became visually detectable through the bandpass filter.

\subsection{Mass spectrometry procedure}
Ion mass spectra were collected using a quadrupole mass spectrometer (Hiden EQP-9 Series 1000). The collector, which used a 50-micron orifice, was positioned at the plasma edge, 0.5 cm radially from the outer edge of the electrodes, and vertically centered between the two electrodes. Ions are drawn through the orifice and directed into the quadrupole by a system of electrodes \citep{howling1994time}; the quadrupole selects the ions by the charge-to-mass ratio, and they are finally collected by a Faraday cup detector. The instrument was set to sweep over a specified mass range of 0-250 amu in 1-amu increments.

Because negative ions are confined within the bulk plasma by the ambipolar electric field, their steady-state flux to the collector is negligible. To release them, the RF power was modulated by a 500 Hz square wave, switching the plasma on and off in 1 ms intervals. Positive ions stream continuously to the collector during the on interval. During each off-interval, the ambipolar field collapses as the remaining positive ions escape, releasing the confined negative ions. 

The 500 Hz modulation signal was fed to the spectrometer's control electronics, and ion counting was synchronized to the cycle. Positive ions were collected over a 1 ms gating window in the plasma-on period, and negative ions over the 1 ms plasma-off window. A dwell time of 50 ms per mass increment was used, meaning that ion counts were accumulated over 50 plasma on-off cycles. 

The instrument's detection efficiency decreases with ion mass due to the combined effects of the Einzel lens transmission, quadrupole transmission, and secondary electron multiplier efficiency. The transmission of the Einzel lens was maximized using a tuning procedure in the spectrometer's software to optimize transmission of ions at the center of the scanning range before measurements were collected. With this tuning, residual non-uniformity across the full range is expected to be small compared to the quadrupole and SEM contributions, and is neglected.

Based on a fit to the manufacturer's transmission curve \citep{HidenAppNote282}, the combined sensitivity relative to mass of the quadrupole transmission and the secondary electron multiplier follow:
\begin{equation}
    \log_{10} \frac{S(M)}{S(1)} = -7.86 \times 10^{-3} \; (M - 1).
\end{equation}

\noindent Raw ion counts were corrected by multiplying by the inverse of this factor at each mass prior to further analysis. 

In the plots presented of the mass spectrum, these corrected spectra were smoothed with a 5-amu Gaussian window to emphasize the long-range periodicity. To calculate the average cluster sizes, the spectra were binned in 18-amu intervals; for negative ions, 17 amu ($\mathrm{OH^-}$) was the dominant bare ion; 18-35 amu was considered $n = 1$, 36-53 amu was $n = 2$, etc.

\subsection{DFT computations}
In DFT, the Gibbs free energy is computed as $\Delta G=\Delta H-T\Delta S$. The total entropy is computed as the sum of the electronic, vibrational, rotational, and translational entropies. Enthalpy $H$ is computed as $H = U + k_B T$. The total energy $U$ is the combination of the energy from the electronic structure, the translational and rotational thermal energies, the zero-temperature vibrational energy, and the finite-temperature vibrational energy due to population of excited vibrational states. These terms are computed from the vibrational modes of the relaxed geometries \citep{Grimme2012,Herzberg1945} using the ORCA software \cite{ORCA}. Initial geometries are taken from Refs.~\cite{Zanuttini2015,Lemke2021}. All geometries were optimized at the DFT/B3LYP-D3/6-311++G** level of theory before the thermochemistry analysis. 

Gibbs free energies $G_{\mathrm{OH^-}}$, $G_{\mathrm{H_2O}}$, and $G_{\mathrm{OH^-(H_2O)_n}}$ were computed at each temperature using ORCA. The formation Gibbs free energy of each cluster relative to its dissociated components is given by Eq. \ref{eq:dG-DFT}, and is corrected from atmospheric pressure to the experimental water partial pressure of 4 Pa via Eq. \ref{eq:pressure-correction}.

For each cluster of $n > 1$, there are multiple possible cluster geometries. Results identify the minimum-energy geometry, as it is the most probable to form.

The relative probability of observing a cluster of size $n$ at temperature $T$ is taken as the Boltzmann factor:
\begin{equation}
    P_n \propto \exp\left(-\frac{\Delta G_n}{k_B T}\right).
\end{equation}

The distribution is normalized such that $\int P_n \, dn = 1$, evaluated numerically in MATLAB via the trapezoidal rule over the computed cluster sizes $n = 1$--$17$. The mean cluster size and standard deviation are then:
\begin{align}
    \langle n \rangle &= \int n \, P_n \, dn, \\
    \sigma_n &= \sqrt{\int n^2 P_n \, dn - \langle n \rangle^2}.
\end{align}

\noindent These are the values compared to CNT and experiment in Fig. 4C.

\section{Derivation of $\Delta G$ using classical nucleation theory}

The change in Gibbs free energy through forming a cluster contains three components:
\begin{equation}
    \Delta G = \Delta G_{vol.} + \Delta G_{surf.} + \Delta G_{elec.},
\end{equation}
representing a volume energy, a surface energy, and electrostatic energy.

The volume energy comes from the change in chemical potential by forming a solid from the vapor phase at constant temperature and pressure. Specifically,
\begin{equation}
    \Delta G_{vol.} = (\mu_s - \mu_v) \Delta N_s,
\end{equation}
where $\mu_s$ and $\mu_v$ are the chemical potential of molecules in the solid or vapor phase, respectively, and $\Delta N_s$ is the change in the number of molecules in the solid phase. For a spherical ice particle, $\Delta N_s$ can be expressed in terms of the ice density:
\begin{equation}
    \Delta N_s = \frac{4 \pi r^3}{3} \frac{\rho_{ice}}{m_{H_2O}}.
\end{equation}
The difference in chemical potential, meanwhile, depends on pressure, and is expressed as:
\begin{equation}
    \mu_s - \mu_v = -k_B T \ln\left(\frac{P}{P_{sat}}\right),
\end{equation}
where $P$ is the pressure of vapor just at the surface of the ice, and $P_{sat}$ is the saturated vapor pressure. 

We note that because of the ion-dipole attraction, vapor molecules are attracted to the ice, and thus the pressure at just above surface of the ice is elevated compared to the pressure far away \cite{lapshin2002vapor, nielsen2011charge}.  The vapor pressure is thus modified as a Boltzmann factor:
\begin{equation}
    P = P_\infty \exp \left( \frac{|p E|}{k_B T} \right) = 
    P_\infty \exp \left( \frac{q_ip}{4 \pi r^2 \epsilon_0 k_B T} \right),
\end{equation}
where $q$ is the ion charge, and $p$ is the permanent dipole moment of water. We here assume that water's permanent dipole moment dominates the interaction and neglect the Langevin polarization.  As shown in \citet{nielsen2011charge} this is a very strong effect.

Combining these equations, the volume Gibbs free energy is given by:
\begin{equation}
    \Delta G_{vol.} =-\frac{4 \pi r^3}{3} \frac{\rho_{ice}}{m_{H_2O}} \; k_B T\ln \left( \frac{P_\infty}{P_{sat}} \right) - \frac{r \rho_{ice}}{3 m_{H_2O}} \frac{q_ip}{\epsilon_0}.
\end{equation}

\noindent Notably, the volume energy is negative when the vapor pressure at the grain surface $P > P_{sat}$.  

The final unknown factor is the saturated vapor pressure. Murphy \& Koop (2005) \cite{murphykoop2005review} provide two expressions for the saturated vapor pressure at the relevant temperatures: one over ice, and one over supercooled liquid water. Since nucleating clusters are generally accepted to be amorphous \citep{kritikos2025can}, consistent with Ostwald's step rule which holds that the least-stable phase forms first before later crystallizing, their properties more closely approximate those of supercooled water than of ice. Accordingly, the Murphy \& Koop expression for vapor pressure over supercooled liquid water is used:

\begin{equation}
\begin{split}
    \ln p_{\text{sat}} =& \; 54.842 - \frac{6763.22}{T} - 4.210 \; \ln T + 3.67\times10^{-4} \; T \\
    &+\tanh\left[0.0415\left(T - 218.8\right)\right]\left(53.878 - \frac{1331.22}{T} - 9.445\ln T + 0.014025 \; T\right)
\end{split}
\end{equation}

where $p_{\text{sat}}$ is in Pascals and $T$ is in Kelvin. For the same reason, the ice density was assumed to be $940 \; \mathrm{kg/m^3}$, the density of amorphous ice \cite{nicolov2024phase}, rather than the lower value of stable crystalline ice.

The surface energy is positive and proportional to the surface area, opposing cluster growth ($\Delta G_{surf.} > 0$). For a spherical cluster,
\begin{equation}
    \Delta G_{surf.} = \gamma \; 4 \pi r^2.
\end{equation}

Tabulated values for the surface tension $\gamma$ in literature assume an infinite sheet of ice; however, the small radius of curvature inflates the surface tension according to the Tolman correction, 

\begin{equation}
    \gamma =  \frac{\gamma_\infty}{1 + 2\delta / r},
\end{equation}

where $\gamma_\infty = 0.109$ N/m is the surface tension of a flat ice surface. The Tolman length $\delta$ is also a matter of debate-- even whether it is positive or negative in value-- we here apply a value of $\delta = -0.4$ \AA, a result of DFT simulations by Ref. \cite{kritikos2025can}. 

The final term accounts for the electric attraction between the condensed water molecules and the central ion. This always serves to decrease the Gibbs free energy, making clustering more energetic favorable. The term is most important for small clusters and results in the downslope at small $n$ seen in Fig. 4A. This is quantified using the dielectric constant of ice, yielding the form \citep{gumbel2009charged,girshick2024nucleation}:

\begin{equation}
    \Delta G_{elec.} = -\frac{q_i^2}{8 \pi \epsilon_0} \left(1 - \frac{1}{\epsilon_r} \right)\left( \frac{1}{r_i} - \frac{1}{r} \right).
\end{equation}

Here, the radius of the central ion is $r_i = 1.1$ \AA, the radius of hydroxide \citep{bellan2022}. The dielectric constant of ice is heavily dependent on temperature; however, as $\epsilon_r >> 1$ for all temperatures, the exact value has negligible impact on the energy. 

Combining the volume, surface, and electric energies, we obtain Eq. 1 in the main text.

\bibliographystyle{aasjournal}
\bibliography{bibliography.bib}

\end{document}